\documentclass[]{emulateapj}

\usepackage{epsfig}

\usepackage{multirow}

\usepackage{rotating}
\usepackage{natbib}
\usepackage{latexsym}
\bibpunct{(}{)}{;}{a}{}{,}

\begin{document}

\newcommand{\ledd}{%
$L_{Edd}$}

\long\def\symbolfootnote[#1]#2{\begingroup%
\def\thefootnote{\fnsymbol{footnote}}\footnote[#1]{#2}\endgroup} 

\def\rem#1{{\bf #1}}
\def\hide#1{}

\title{On the cooling tails of thermonuclear X-ray bursts: the IGR J17480--2446 link}

\author{M. Linares\altaffilmark{1,2}, D. Chakrabarty\altaffilmark{1}, M. van der Klis\altaffilmark{3}}


\submitted{Accepted for publication in The Astrophysical Journal Letters}

\altaffiltext{1}{Massachusetts Institute of Technology - Kavli Institute for Astrophysics and Space Research, Cambridge, MA 02139, USA}

\altaffiltext{2}{Rubicon Fellow}

\altaffiltext{3}{Astronomical Institute ``Anton Pannekoek'', University of Amsterdam, PO BOX 94249, 1090 GE, Amsterdam, Netherlands}



\begin{abstract}

The neutron star transient and 11~Hz X-ray pulsar IGR~J17480--2446,
recently discovered in the globular cluster Terzan~5, showed
unprecedented bursting activity during its 2010 October--November
outburst. We analyzed all X-ray bursts detected with the {\it Rossi
  X-ray Timing Explorer} and find strong evidence that they all have a
thermonuclear origin, despite the fact that many do not show the
canonical spectral softening along the decay imprinted on type~I X-ray
bursts by the cooling of the neutron star photosphere. We show that
the persistent-to-burst power ratio is fully consistent with the
accretion-to-thermonuclear efficiency ratio along the whole outburst,
as is typical for type~I X-ray bursts. The burst energy, peak
luminosity and daily-averaged spectral profiles all evolve smoothly
throughout the outburst, in parallel with the persistent (non-burst)
luminosity. We also find that the peak burst to persistent luminosity
ratio determines whether or not cooling is present in the bursts from
IGR~J17480--2446, and argue that the apparent lack of cooling is due
to the ``non-cooling'' bursts having both a lower peak temperature and
a higher non-burst (persistent) emission. We conclude that the
detection of cooling along the decay is a sufficient, but not a
necessary condition to identify an X-ray burst as
thermonuclear. Finally, we compare these findings with X-ray bursts
from other rapidly accreting neutron stars.

\end{abstract}

\maketitle

\section{Introduction}
\label{sec:intro}

Since soon after their discovery \citep{Grindlay76}, the distinctive
property of thermonuclear bursts from accreting neutron stars (NSs)
has been the presence of a thermal spectrum with temperature
decreasing along the burst decay
\citep{Swank77,Hoffman77,Hoffman78}. Such ``cooling tails'' are
attributed to the NS photosphere cooling down after the fast injection
of heat from nuclear reactions, deeper in the ocean. An X-ray burst
showing a cooling tail is {\it classified} as a type~I X-ray burst and
{\it identified} as thermonuclear. On the other hand, a different kind
of X-ray bursts, with integrated energies similar to or greater than
the accretion energy radiated between bursts, was classified as type
II and identified with spasmodic accretion events onto a NS
\citep[][and references therein]{Lewin93}. Thousands of type~I X-ray
bursts have been observed from more than 90 NS low-mass X-ray binaries
(NS-LMXBs) to date \citep[e.g.][]{Cornelisse03,Galloway08}. Their
  thermonuclear nature bears little or no doubt and models have met
  with considerable success in reproducing many of their properties
  \citep[e.g.,][and references therein]{Strohmayer06}. Type II bursts
have only been detected with confidence from two sources \citep[the
  ``rapid burster'' MXB~1730--335, and the ``bursting pulsar''
  GRO~J1744--28][respectively]{Lewin93,Kouveliotou96}. This, and the
fact that bursts from these two sources showed different spectral
properties, has led to a somewhat ambiguous definition of type~II
bursts. In some cases type~II bursts have been implicitly defined as
X-ray bursts lacking cooling tails
\citep[e.g.,][]{Kuulkers02,Galloway08} while in other cases they are
simply described as ``spasmodic accretion events'', a definition that
lacks a strict observational criterion.

An X-ray transient in the direction of the globular cluster Terzan~5
was detected with {\it INTEGRAL} on 2010-10-10 \citep{Bordas10}
and showed a type~I X-ray burst one day later
\citep{Chenevez10}. Pulsations and burst oscillations at $\sim$11~Hz
were discovered with {\it RXTE} on 2010-10-13
\citep[][respectively]{Strohmayer10,Altamirano10a}. This, together
with the {\it Chandra} localization \citep{Pooley10}, showed that the
source was a new transient NS-LMXB, which was named IGR~J17480--2446
\citep[previously identified as a quiescent LMXB candidate by][labeled
  CX25 or CXOGlb~J174804.8--244648]{Heinke06}. We refer to it
hereafter as T5X2, as this is the second bright X-ray source
discovered in Terzan~5. T5X2 is in a 21~hr orbit \citep{Papitto11} and
contains the slowest rotating NS among the type~I X-ray burst sources
with known spin. \citet{Linares10c} discovered millihertz
quasi-periodic oscillations during the peak of the outburst \citep[see
  also][]{Linares11}.

It was suggested that some of the bursts from T5X2 were type~II, based
on the non-detection of cooling tails \citep{Galloway10}.
\citet{Chakraborty11} have suggested a thermonuclear origin of the
bursts from T5X2, as initially argued by \citet{Linares10c}, although
they do not present a detailed burst spectral analysis nor do they
discuss in detail the causes and implications of non-cooling
thermonuclear bursts (we refer to the bursts where cooling is not
detectable as ``non-cooling'' bursts; see Sec.~\ref{sec:results} \&
\ref{sec:discussion} for details). We present in this Letter strong
evidence that all the bursts observed from T5X2 in 2010 were
thermonuclear, even those that did not show a cooling tail and
therefore cannot be classified as type~I X-ray bursts. We show that
the burst thermal profile evolves gradually as the persistent
luminosity changes along the outburst, and as a consequence
bursts change from type~I to non-cooling and back to type
I. Furthermore, we show that the presence/absence of cooling along
the tail is determined by the ratio between peak burst and persistent
luminosity. We put forward an explanation for the lack of cooling,
discuss the implications of these results for the identification of
thermonuclear bursts and compare them with other NS-LMXBs that show
X-ray bursts at high persistent luminosities ($L_{pers}$) and inferred
mass accretion rates ($\dot{M}$).

\begin{figure}[ht]
\centering
  \resizebox{0.9\columnwidth}{!}{\rotatebox{0}{\includegraphics[]{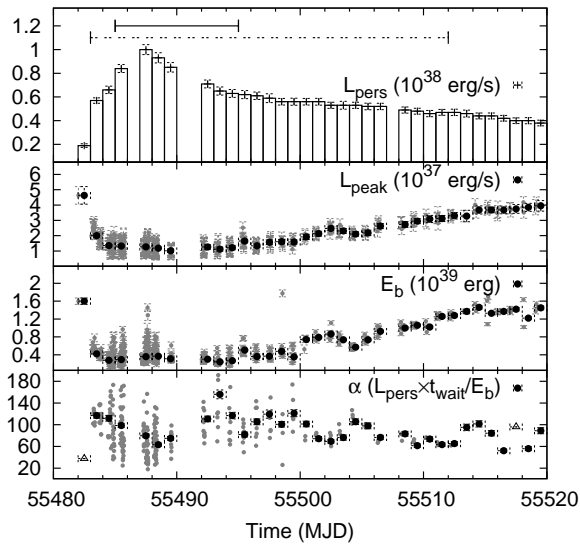}}}
  \caption{From top to bottom: evolution along the T5X2 outburst of {\it i)} persistent (accretion) 2--50~keV luminosity; {\it ii)} bolometric peak burst luminosity; {\it iii)} bolometric radiated burst energy and {\it iv)} alpha parameter (accretion-to-burst energy ratio). All energies and luminosities use a distance of 6.3~kpc (Sec.~\ref{sec:data}). Grey and black circles show individual burst measurements and daily averages, respectively. Open triangles show alpha daily averages based on one single burst, which we consider lower limits. Horizontal lines along the top axis show the time intervals when no cooling is detected in individual bursts (dashed line) and daily-averaged burst spectral profiles (solid line). MJD~55480 is 2010-10-11.}
    \label{fig:ob}
\end{figure}

\section{Data analysis}
\label{sec:data}

We analyzed all 46 {\it Rossi X-ray Timing Explorer (RXTE)}
observations of T5X2 taken between 2010-10-13 and 2010-11-19
(MJDs~55482--55519; proposal-target number 95437-01), using data
from all available proportional counter units (PCUs). After that date
the source was not observable with {\it RXTE} due to Solar
constraints\footnote{{\it RXTE} and {\it MAXI} observations did not
  detect T5X2 after the Sun-constrained period ended, on 2011-01-17
  and 2010-12-27, respectively.}. We visually inspected 2~s time
resolution 2--30~keV light curves throughout the outburst to search
for X-ray bursts. We used Event data (or Good-Xenon when available) to
perform time-resolved spectroscopy of all bursts, extracting dead time
corrected spectra in 2~s time bins and using a $\sim$100~s-long pre-
or post-burst interval to extract the non-burst (persistent emission
plus background) spectrum. We fitted the resulting spectra within
Xspec (v 12.6.0q) using a simple blackbody model and fixing the
absorbing column density to 1.2$\times$10$^{22}$ cm$^{-2}$
\citep{Heinke06}, after adding a 1\% systematic error to all channels
and grouping them to a minimum of 15 counts per channel. 

We use throughout this work a distance of 6.3~kpc to Terzan~5, the
highest value reported by \citet{Ortolani07} from {\it HST}
photometry, which is consistent with the independent measurement of
\citet{Galloway08} based on photospheric radius expansion bursts from
a burster in the same globular cluster. We note that recent
measurements of the distance to Terzan~5 span the range 4.6--8.7~kpc
\citep{Cohn02,Ortolani07} and therefore the luminosities and energies
presented herein have a systematic uncertainty of a factor $\sim$2
(factor $\sim \sqrt{2}$ for blackbody radii). We stress, however, that
the distance uncertainty does not affect our conclusions as the two
main parameters we use ($\alpha$ and $\beta$, as defined below) do not
depend on the distance to Terzan 5.

We integrated the bolometric luminosity along each burst to find the
total energy radiated, $E_b$, and defined the ratio of accretion to
nuclear energy as $\alpha$$\equiv$$L_{pers}$$\times$$t_{wait}$ /
$E_b$, where $L_{pers}$ is the persistent (accretion) luminosity (see
below) and $t_{wait}$ is the time since the previous burst or wait
time, available when no data gaps are present before a given
burst. The brighter the persistent (accretion) flux is, the fainter
bursts become: between MJD~55482 and 55488 (2010-10-13~and~2010-10-19)
the net peak burst flux decreases from $\sim$75\% to $\sim$10\% of the
total flux, while the non-burst count rate increases from $\sim$250
c~s$^{-1}$~PCU$^{-1}$ to $\sim$1600
c~s$^{-1}$~PCU$^{-1}$. Consequently, spectroscopy of individual bursts
becomes limited by the low signal-to-noise.

In order to study the burst spectral properties along the
outburst, we divided each burst in six intervals, with the following
time ranges in seconds relative to the peak burst time: $-$6$\pm$4,
0$\pm$2, 6$\pm$4, 15$\pm$5, 25$\pm$5 and 35$\pm$5.  After rejecting
individual 2s time bins with ill-constrained spectral parameters
(error larger than value), we find the daily-averaged spectral
parameters (blackbody radius, temperature and luminosity) in each of
these six time intervals. Furthermore, in order to quantify the
  amount of cooling we define the temperature drop $\Delta
  T$$\equiv$$T_{peak}$-$T_{15}$, where $T_{peak}$ and $T_{15}$ are the
  daily-averaged temperatures in the time intervals 0$\pm$2 and
  15$\pm$5~s, respectively. To study the persistent emission we
extracted one dead time corrected spectrum per observation from
Standard~2 data, excluding all bursts and subtracting the background
spectrum given by the bright source background model. We fitted the
resulting spectra with a model sum of disk blackbody, power law and
Gaussian line with energy fixed at 6.5~keV, correcting for absorption
as above. We obtain the 2--50~keV persistent luminosity ($L_{pers}$)
from the best fit model. We do not apply bolometric correction
  when calculating $L_{pers}$, which would increase its value (and
  $\alpha$) by about a factor of 2 (see, e.g., \citealt{intZand07}).


\section{Results}
\label{sec:results}

\begin{figure}[t!]
\centering
  \resizebox{0.65\columnwidth}{!}{\rotatebox{0}{\includegraphics[]{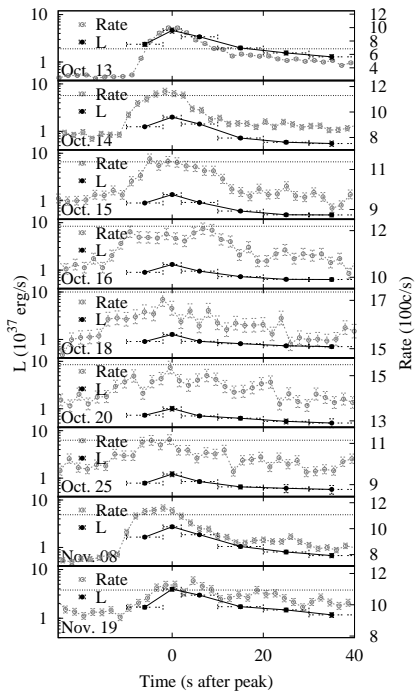}}}
  \caption{Burst light curves for nine selected dates. Shown are: count rate light curve of one representative burst (grey circles; PCU2; $\sim$2--60~keV; not background subtracted) and daily-averaged bolometric burst luminosity (filled black circles; dotted lines show persistent luminosity, $L_{pers}$).}
    \label{fig:bspec}
\end{figure}

\begin{figure}[t!]
\centering
  \resizebox{0.65\columnwidth}{!}{\rotatebox{0}{\includegraphics[]{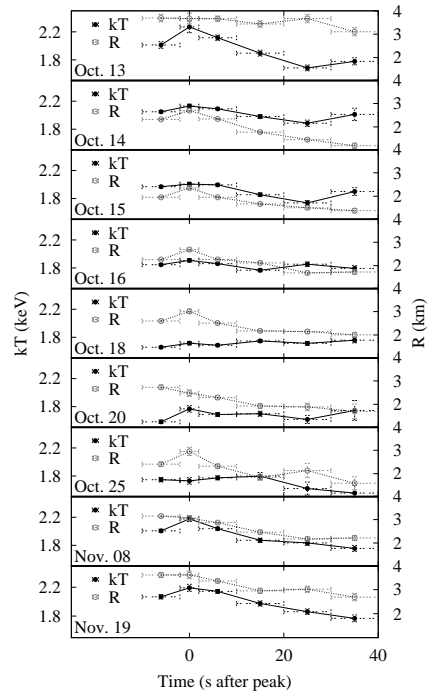}}}
  \caption{Burst spectral properties in six time bins for nine selected dates. Shown are: daily-averaged blackbody temperature (filled black circles) and blackbody radius (grey circles; for a distance of 6.3~kpc, see Sec.~\ref{sec:data}).}
    \label{fig:bspec2}
\end{figure}

\begin{figure}[h!]
  \begin{center}
  \resizebox{0.85\columnwidth}{!}{\rotatebox{0}{\includegraphics[]{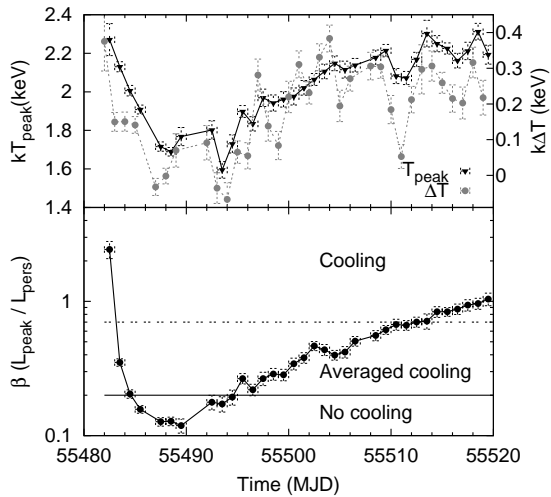}}}
  \caption{{\it Top:} Evolution along the T5X2 outburst of the daily-averaged peak blackbody temperature (T$_{peak}$) and temperature drop ($\Delta T$; Sec.~\ref{sec:data}). {\it Bottom:} Evolution of the ratio $\beta$ between peak-burst and persistent luminosities. Bursts are canonical type~I X-ray bursts when $\beta$$\gtrsim$0.7. Bursts with $\beta$$\lesssim$0.2 do not show any spectral cooling. In the range 0.7$\gtrsim$$\beta$$\gtrsim$0.2 individual bursts show no cooling tails, but we are able to measure significant cooling in the daily-averaged spectral profiles.}
    \label{fig:cool}
 \end{center}
\end{figure}

We find a total of 373 X-ray bursts between 2010-10-13 and
2010-11-19 (MJDs~55482--55519). Figure~\ref{fig:ob} summarizes the
joint evolution of burst energetics and persistent luminosity: while
$L_{pers}$ increases bursts become fainter and less energetic, whereas
burst energy and luminosity increase along the outburst decay, when
$L_{pers}$ drops (see also Figure~\ref{fig:bspec}). The first burst
detected by {\it RXTE} (on 2010-10-13; MJD~55482) and all (20) bursts
detected after 2010-11-11 (MJD~55511) show clear cooling tails and can
be unequivocally classified as type~I X-ray bursts \citep[see
  also][]{Galloway10}. By averaging burst profiles during each day
(see Figure~\ref{fig:bspec2}; Sec.~\ref{sec:data}), we also find
evidence of cooling until 2010-10-15 (MJD~55484) and after 2010-10-25
(MJD~55494). Between these dates (i.e., between MJDs~55485--55494) we
do not detect cooling along the burst tail, in neither individual
bursts nor daily averaged spectral profiles. This can be seen
quantitatively in the temperature drop, $\Delta T$, which is
consistent with zero between MJDs~55485--55494 (Figure~\ref{fig:cool},
top).

Both burst peak luminosity ($L_{peak}$) and burst energy ($E_b$) are
anticorrelated with $L_{pers}$ and follow a smooth evolution, even
when cooling becomes first undetectable and then detectable again. The
presence of canonical type~I X-ray bursts when $L_{pers}$ is
relatively low, together with their smooth evolution into faint
``non-cooling'' bursts, give strong evidence that {\it all} X-ray
bursts from T5X2 are thermonuclear. The $\alpha$ values span an
approximately constant range along the whole outburst, between
$\sim$20 and $\sim$200, fully consistent with a thermonuclear origin
of the bursts \citep[e.g.][]{Lewin93}. The gradual evolution of the
burst rate \citep{Linares10c} and the presence of burst oscillations
in both cooling and non-cooling bursts \citep{Cavecchi11} also suggest
a common origin for all bursts. Furthermore, the daily-averaged
spectral profiles (Fig.~\ref{fig:bspec2}) evolve smoothly along the
outburst: the peak blackbody temperature decreases from $\sim$2.3~keV
to $\sim$1.7~keV between the first observation (2010-10-13; MJD~55482)
and the outburst maximum (2010-10-18; MJD~55487), and later increases
up to $\sim$ 2.2~keV on 2010-11-19 (MJD~55519; see also
Figure~\ref{fig:cool}, top panel). Such smooth spectral evolution
further strengthens the thermonuclear identification of all bursts
from T5X2. The apparent emitting area we find is remarkably low,
  with peak daily-averaged blackbody radii between 2 and 4 km
  throughout the whole outburst \citep[Fig.~\ref{fig:bspec2}; see also
    discussion in][]{Cavecchi11}.

\begin{figure*}[ht]
  \begin{center}
  \resizebox{1.7\columnwidth}{!}{\rotatebox{-90}{\includegraphics[]{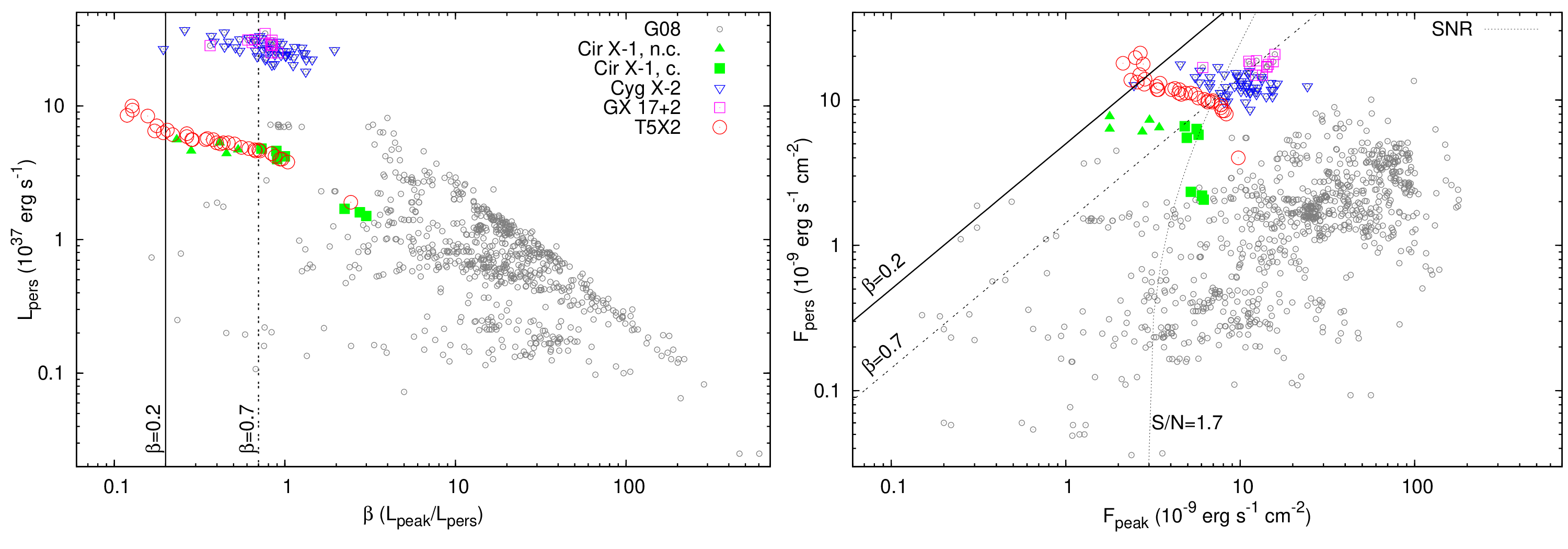}}}
\vspace{-0.4cm}  \caption{{\it Left:} Persistent luminosity ($L_{pers}$) versus $\beta$, for different bursters. {\it Right:} Persistent flux versus burst peak flux. Small gray circles show all bursts from the {\it RXTE}-MIT type~I X-ray burst catalog \citep[][$L_{pers}$ in the 2.5--25~keV band estimated from their parameter $\gamma$]{Galloway08}. The burst daily averages from T5X2 are shown in red filled circles (this work; $L_{pers}$ in the 2--50~keV band). Interestingly, cooling bursts (green filled squares) and non-cooling bursts (green filled triangles) from Cir~X-1 are clearly differentiated by the same $\beta$=0.7 threshold (dashed line) that divides individual cooling and non-cooling bursts in T5X2 (see Fig.~\ref{fig:cool} and Sec.~\ref{sec:results}; $L_{pers}$ for Cir~X-1 in the 0.5--50~keV band). A signal-to-noise ratio criterion, (dotted line; Section~\ref{sec:discussion}), does not distinguish between cooling and non-cooling bursts from both T5X2 and Cir~X-1. Other two high-$\dot{M}$ sources that have shown non-cooling bursts are indicated, GX~17+2 (open pink squares) and Cyg~X-2 (open blue circles), and lie on the $\beta$=0.7 boundary.}
    \label{fig:galloway}
 \end{center}
\end{figure*}

We find that the ratio of peak burst to persistent luminosity,
$\beta$$\equiv$$L_{peak}$/$L_{pers}$, determines whether or not
cooling is manifest in the T5X2 bursts (see
Figure~\ref{fig:cool}, bottom panel). For $\beta$$\gtrsim$0.7
individual bursts show cooling along the tail. In the range
0.7$\gtrsim$$\beta$$\gtrsim$0.2, we do not measure cooling in
individual bursts, but we do find significant cooling in the
daily-averaged temperature profiles. For $\beta$$\lesssim$0.2 the
imprints of cooling vanish also in the daily averages. These
$\beta$=[0.7,0.2] critical values translate into the time ranges shown
in the top panel of Figure~\ref{fig:ob} and correspond to
  $L_{pers}$$\simeq$[4.7,6.5]$\times$10$^{37}$~erg~s${-1}$ and
  $L_{peak}$$\simeq$[3.3,1.3]$\times$10$^{37}$~erg~s${-1}$. Comparison with
  sources with higher $L_{pers}$ that show both cooling and
  non-cooling bursts indicates that the presence or absence of
  measurable cooling along the burst decay does not depend solely on
  $L_{peak}$ or $L_{pers}$ but on their ratio $\beta$, as we discuss
  in detail in the following Section.

\section{Discussion}
\label{sec:discussion}

Our work shows that the absence of measurable cooling along the tail
does not rule out the thermonuclear origin of an X-ray burst, and that
the type~II burst classification cannot merely rely on the lack of
spectral cooling. This implies that the detection of cooling along the
decay is a sufficient, but not a necessary condition to identify an
X-ray burst as thermonuclear. We have characterized the burst spectral
properties along the whole outburst, and shown that the
$\beta$$\equiv$$L_{peak}$/$L_{pers}$ ratio determines whether or not
cooling is measurable in the bursts from T5X2 (Fig.~\ref{fig:cool};
Sec.~\ref{sec:results}). Individual bursts from T5X2 with
$\beta$$\lesssim$0.7 do not show significant cooling along the
decay. The main reason why non-cooling bursts had not been firmly
identified as thermonuclear until now is that the vast majority of
thermonuclear bursts have $\beta$$>$0.7
\citep[Figure~\ref{fig:galloway};][]{Galloway08} and therefore show
clear cooling tails, while only a handful of bursts with $\beta$$<$0.7
had been observed to date. We stress that any X-ray burst detected
with $\beta$$>$0.7 should still show cooling along the tail in order
for it to be identified as thermonuclear.

The gradual disappearance of spectral cooling can be attributed to a
combination of two factors. Firstly, the sensitivity to spectral
cooling decreases as persistent emission rises. Our spectral fits show
that blackbody area and temperature are increasingly correlated when
the signal-to-noise ratio decreases. Using a single parameter model
(keeping the blackbody temperature or emitting area fixed throughout
the burst) gives statistically acceptable fits to the faint bursts,
indicating that the temperature and emitting area cannot be
simultaneously determined by performing standard time-resolved
spectroscopy of individual faint bursts. By assuming a constant
emitting area and fixing it to the average value found for the first
and brightest burst we can recover a temperature decay along the tail
of the faintest bursts. Alternatively, the temperature could remain
approximately constant throughout the bursts if the burning area
changes size. Such ``single-parameter approach'', however, has
obviously severe limitations as it relies on an arbitrary choice of
constant emitting area or temperature.

On the other hand, we find that as $L_{pers}$ (and hence the inferred
mass accretion rate, $\dot{M}$) increases bursts become colder and
less energetic, probably due to a lower ignition column depth
resulting in less mass being burned. Less energetic bursts result into
lower peak temperatures and a less pronounced temperature
decay. Moreover, the higher $\dot{M}$ could also lead to a hotter NS
photosphere {\it between} bursts as more accretion energy is being
released on the NS surface, which would rise the ``baseline
temperature'', again smearing out the burst cooling tail
\citep[see][for further discussion]{Zand09}. We thus conclude
that the lack of measurable cooling is due to i) a loss of sensitivity
when persistent emission increases and ii) a decrease in burst
temperature (and luminosity) when $\dot{M}$ increases. It is likely
that the NS photosphere cools down during the tails of the
``non-cooling'' bursts, but by an amount below our detection
limit.

\citet{Paradijs86} pointed out that if a significant fraction of the
persistent flux is thermally emitted from the NS surface the
``standard'' burst spectral analysis (which subtracts pre-burst
emission) can underestimate the blackbody radius and overestimate its
temperature, in particular when the burst flux is low. The reason is
that the difference between two blackbody spectra (one powered by
accretion and nuclear energy minus one only powered by accretion) does
not have a Planckian distribution \citep{Paradijs86}. If such
systematic effect is present in the faint, high-$\dot{M}$ bursts and
increases the apparent temperature with decreasing burst flux, that
could also reduce the amount of cooling observed. Hence the apparent
gradual decrease in the amount of cooling ($\Delta T$) observed when
$\beta$ decreases (Fig.~\ref{fig:cool}) could be partially caused by
the standard burst analysis overestimating the blackbody temperature
in a gradually increasing fraction of the burst. We note, however,
that the evolution of $kT_{peak}$ is robust, i.e., bursts become
intrinsically colder and less luminous when $L_{pers}$ increases.



It is interesting to note that non-cooling bursts similar to the T5X2
ones presented herein were detected from Circinus~X-1 (Cir~X-1) in May
2010, and tentatively identified as thermonuclear bursts
\citep{Linares10d}. The ``early bursts'' reported by
\citet{Linares10d} did not show cooling tails whereas later bursts
became canonical type~I X-ray bursts, a behavior reminiscent of T5X2
\citep[see also][]{Tennant86a}. We have calculated $\beta$ for all the
2010 bursts from Cir~X-1 (see Figure~\ref{fig:galloway}). The same
critical value of $\beta$ that we find for T5X2, blindly applied to
Cir~X-1, clearly differentiates between the cooling ($\beta$$>$0.7)
and non-cooling ($\beta$$<$0.7) bursts. We show in
Figure~\ref{fig:galloway} (left) the $\beta$ values plotted versus
$L_{pers}$ for all bursts reported by \citet{Galloway08}, together
with Cir~X-1 and T5X2. Furthermore, defining the signal-to-noise
ratio as S/N$\equiv F_{peak} / \sqrt{F_{pers} + F_{peak}}$ (where
$F_{peak}$ and $F_{pers}$ are the peak-burst and persistent fluxes,
respectively) we explore whether or not S/N determines the presence of
measurable cooling tails. Figure~\ref{fig:galloway} (right) shows that
a simple S/N criterion fails to distinguish the type~I and non-cooling
bursts from both T5X2 and Cir~X-1, indicating that the lack of
measurable cooling is not merely due to low statistics and that
$\beta$ provides a better criterion to identify non-cooling bursts.

To our knowledge two other NS-LMXBs, GX~17+2 and Cyg~X-2, have shown
non-cooling X-ray bursts that have not been conclusively identified
\citep[although a thermonuclear origin has been
suggested;][]{Kahn84,Sztajno86,Kuulkers95,Kuulkers02,Galloway08}. Given
that they also cross the $\beta$=0.7 boundary
(Fig.~\ref{fig:galloway}) it is not surprising that these two sources
show both cooling and non-cooling bursts. The cooling bursts from
GX~17+2 and Cyg~X-2 occurred when $L_{pers}$ and $L_{peak}$ were
considerably ($\sim$5 times when $\beta$$\simeq$0.7) higher than the
T5X2 case presented here (Fig.~\ref{fig:galloway}), which strongly
suggests that it is $\beta$ and not simply $L_{pers}$ or $L_{peak}$
that determines whether or not cooling tails are
observed. Figure~\ref{fig:galloway} (right) displays a large number of
cooling, canonical type~I X-ray bursts with the same peak fluxes as
the non-cooling bursts presented here, showing that a non-cooling
burst can not be predicted from its net peak flux alone. In all four
sources discussed above the accretion luminosity at the time of the
bursts was relatively high
($L_{pers}$$\gtrsim$5$\times$10$^{37}$~erg~s$^{-1}$) compared to that
of most bursters, and all four sources have shown Z source
behavior. This suggests that non-cooling thermonuclear bursts are an
exclusive feature of high-$\dot{M}$ states of NS-LMXBs. The case of
T5X2 presented herein, however, remains exceptional for its extremely
high burst rate \citep{Linares11} and for the unprecedented smooth
evolution between canonical type~I X-ray bursts and non-cooling
high-$\dot{M}$ bursts. It also opens the prospect of a ``hidden'',
previously unrecognized, population of non-cooling bursts from rapidly
accreting NSs.\\

\textbf{Acknowledgments:} 

We thank Diego Altamirano, Andrew Cumming, Jean in 't Zand, Laurens
Keek and Walter Lewin for insightful discussions and comments on the
manuscript, and Michael Zamfir for sharing his results prior to
publication. ML acknowledges support from the Netherlands Organization
for Scientific Research (NWO) Rubicon fellowship.\\


\end{document}